# Bound and free waves in non-collinear second harmonic generation

M.C.Larciprete, [1,*] F.A.Bovino,[2] A. Belardini,[1] C. Sibilia,[1] and M. Bertolotti,[1]

[1]*Dipartimento di Energetica, Università di Roma La sapienza, Via A.Scarpa 16 00161 Roma, Italy*
[2] *Quantum Optics Lab. , Elsag-Datamat Via Puccini 2 Genova, Italy*
[*]*Corresponding author: mariacristina.larciprete@uniroma1.it*

**Abstract:** We analyze the relationship between the bound and the free waves in the noncollinear SHG scheme, along with the vectorial conservation law for the different components arising when there are two pump beams impinging on the sample with two different incidence angles. The generated power is systematically investigated, by varying the polarization state of both fundamental beams, while absorption is included via the Herman and Hayden correction terms. The theoretical simulations, obtained for samples which are some coherence length thick show that the resulting polarization mapping is an useful tool to put in evidence the interference between bound and free waves, as well as the effect of absorption on the interference pattern.

## 1. Introduction.

The rotational Maker fringes technique is a well established method [1] introduced by Maker and coworkers to characterize the second order nonlinear optical tensor of a given material. This method consists in measuring the second harmonic (SH) signal as a function of the fundamental beam incidence angle for a given polarization state of both the input fundamental and SH output beam. Leaving aside third and higher harmonics, the electric field tuned at $\omega_1$ induces in the nonlinear material a polarization composed by two waves tuned at $2\omega_1$. Given the refractive index dispersion, the so-called "bound" and the "free" wave experience $n(\omega_1)$ and $n(2\omega_1)$, respectively, and generally travel at different velocities. The existence of these two waves, simply obtained as solutions of Maxwell's equations [2], is at the basis of this well established technique. When the laser pulses coherence length is longer than sample thickness, the bound and free waves give rise to interference fringes which appear in the generated harmonic power as the slab is rotated. The second order susceptibility of the investigated crystal is thus calculated from the angle dependence of the harmonic power [3,4]. The bound wave is also referred as the *phase-locked* wave, in order to point out that it is located under the pump pulse and is dragged at the pump's group velocity [5].

Since the introduction of many organic nonlinear materials, it was no more possible to neglect the effect of absorption on both amplitude and shape of the Maker fringes curves. Thus the Maker theory was revised by Herman and Hayden [6] in order to take into account the absorption of the nonlinear material. In this way it is possible to reconstruct the angular behavior of SH signal also when absorption changes the absolute value and the shape of the fringes. In the strong absorption regime, in fact, only the bound wave survives in the nonlinear medium, as experimentally shown by several authors [7,8,9].

Noncollinear SHG is an important technique that provides new capabilities in the characterization of nonlinear materials, thin films in particular. The technique was firstly demonstrated by Muenchausen [10] and Provencher [11]. Later on, Figliozzi [12] and Cattaneo [13] have shown that the technique allows the bulk and surface responses to be addressed. Furthermore, Cattaneo has also demonstrated that the technique is very useful in surface and thin-film characterization [14,15]. More recently, a quasi-collinear scheme was employed to analyze the polarization properties of SHG from SBN crystals with disordered ferroelectric domains [16]. We employed noncollinear second harmonic generation, to develop a method based on the simultaneously variation of the polarization state of both fundamental beams, at a fixed incidence angle [17]. This method permits to visualize important crystalline characteristics of nonlinear crystal or films, and allows to address all the different non-zero components of the nonlinear optical tensor [18]. The generated signal can be thus represented as a function of the polarization states of both pump beams. The result is a *polarization map* whose pattern is characteristic of the investigated crystalline structure. This method offers the possibility to evaluate the ratio between the different non-zero elements of the nonlinear optical tensor, or the evaluation of the absolute values of the non-zero terms of the nonlinear optical tensor, without requiring sample rotation. As a result, it is extremely interesting for those conditions where the generated signal would be strongly affected by sample rotation angle, i.e. for samples which are some coherence lengths thick, when using short laser pulses, of for nano-patterned samples. With respect to the given examples, this method of polarization scan allows the characterization of the nonlinear optical tensor elements without varying the experimental conditions.

In this work, we introduce the effect of absorption in the analytical expression for the noncollinear generated SH power, by including the Herman and Hayden correction terms. We will focus our attention on the relationship between the bound and the free waves in the noncollinear SHG scheme, along with the vectorial conservation law for the different components arising when there are two pump beams impinging on the sample with two different incidence angles. As examples, we present the theoretical simulations, obtained for two different crystalline structures, some coherence length thick. The examples show that this method, which doesn't require sample rotation, is also an useful tool to put in evidence the interference between bound and free waves arising from the noncollinear process, as well as the effect of absorption on the resulting interference pattern.

## 2. Wavevector conservation in noncollinear second harmonic generation.

In noncollinear second harmonic generation, two beams of the same intensity are sent to intersect onto the surface of a nonlinear material. The polarization of both beams can be varied systematically, i.e. with two identical rotating half wave plates. Under conditions of wave-plane approximation, we studied the geometrical configuration shown in Fig.1.

We consider two pump beams, tuned at $\omega_1 = \omega_2 = \omega$, having two different incidence angles, with respect to surface normal, $\alpha_1$ and $\alpha_2$, and different polarization state, $\phi_1$ and $\phi_2$ (defined with respect to the y-z plane), respectively. The internal propagation angles of the fundamental beams, $\alpha'_1$ and $\alpha'_2$ are calculated via Snell's law in anisotropic crystal [19]. Given the conservation of the tangential component of momentum at a boundary [20], we found for the wavevectors the situation depicted in Fig.1, once the two external wavevectors, $k_{i1}^\omega = 2\pi/\lambda$ and $k_{i2}^\omega = 2\pi/\lambda$, are incident, from air, onto sample surface. Each of these vectors generates a refracted vector, given by $k_1^\omega = 2\pi n_\omega(\alpha'_1)/\lambda$ and $k_2^\omega = 2\pi n_\omega(\alpha'_2)/\lambda$, being their tangential components preserved.

Inside the nonlinear material, the second order nonlinear optical tensor $\chi^{(2)}$ produces different components of the nonlinear optical polarization, in both collinear and noncollinear directions, being $P^{(2)} = \chi^{(2)} : E^2_{\omega,1} + \chi^{(2)} : E^2_{\omega,2} + \chi^{(2)} : E_{\omega,1} E_{\omega,2}$ [21]. Considering the collinear processes, we find the bound and free wavectors corresponding to the waves tuned at $2\omega_1$ and $2\omega_2$, respectively. The two bound waves, having wavevectors $k_{1,B}^{2\omega} = 4\pi n_\omega(\alpha'_{1,B})/\lambda$ and $k_{2,B}^{2\omega} = 4\pi n_\omega(\alpha'_{2,B})/\lambda$, are collinear with the corresponding pumps and travel at their same velocities. The two propagation angles $\alpha'_{1,B}$ and $\alpha'_{2,B}$, are equal to $\alpha'_1$ and $\alpha'_2$, respectively, and depend on the polarization state of the corresponding pump beam but not on the chosen polarization state of the SH. On the other side, the two free waves experience higher refractive indices, due to material dispersion, thus their wavevectors, $k_{1,F}^{2\omega} = 4\pi n_{2\omega}(\alpha'_{1,F})/\lambda$ and $k_{2,F}^{2\omega} = 4\pi n_{2\omega}(\alpha'_{2,F})/\lambda$, are slightly more refracted towards the surface normal. The resulting propagation angles $\alpha'_{1,F}$ and $\alpha'_{2,F}$, are dependent only on the polarization state of the SH beam.

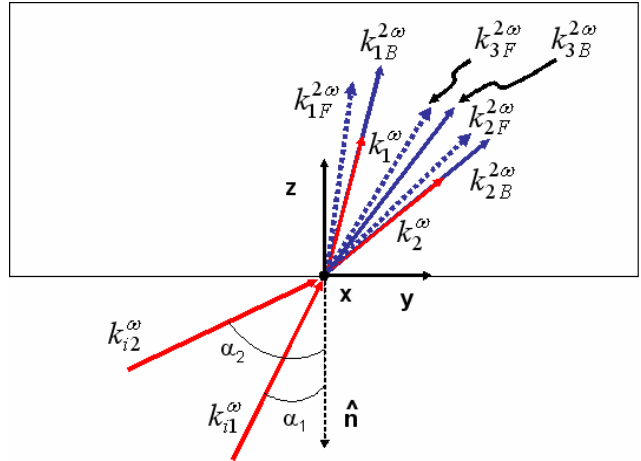

Fig. 1. Scheme of noncollinear second harmonic generation.

As an example, we chose two beams impinging onto Silicon carbide (SiC), with $\alpha_1 = 26°$ and $\alpha_2 = 44°$, i.e. the two pump beams have an aperture angle of 18° with respect to each other. The fundamental beam is tuned at 830 nm and the extraordinary and ordinary refractive index are assumed to be $n_\omega^e = 2.64$ and $n_\omega^o = 2.60$ at the fundamental beam frequency and $n_{2\omega}^e = 2.80$ and $n_{2\omega}^o = 2.75$ at the SH frequency [22]. In Fig.2 we report the propagation angles for the resulting collinear bound waves, $\alpha'_{1,B}$ and $\alpha'_{2,B}$ as a function of polarization state of the corresponding pump beam.

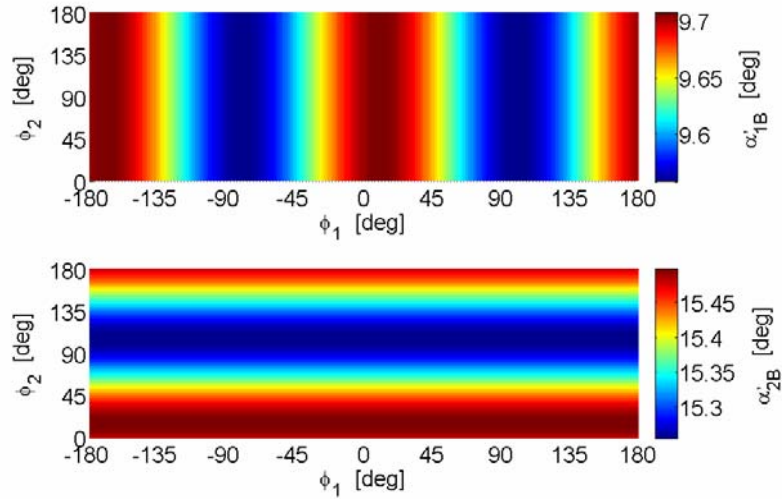

Fig. 2. Propagation angles for the collinear bound waves, $\alpha'_{1,B}$ and $\alpha'_{2,B}$ as a function of polarization state of the corresponding pump beam.

Under the same conditions we find the propagation angles for the two free waves to be fixed at a constant value of $\alpha'_{1,F}$ =9.16° for a $\hat{p}$-polarized SH (or $\alpha'_{1,F}$ =9.17° for an $\hat{s}$-polarized SH), and $\alpha'_{2,F}$ =14.61° for a $\hat{p}$-polarized SH (or $\alpha'_{2,F}$ =14.63° for an $\hat{s}$-polarized SH).

The direction of the noncollinear signal, tuned at ω1+ω2=2ω, is approximately bisecting the angle between the two pump beams, whatever the sample rotation angle. The wavevectors associated with the bound and the free waves arising from the noncollinear process, are indicated with $k_{3,B}^{2\omega}$ and $k_{3,F}^{2\omega}$ respectively, and lay between $k_1^\omega$ and $k_2^\omega$. While both these waves arise in the region in which the two fundamental waves overlap, it's worth noting that the noncollinear bound wave disappear if the two fundamental waves separate each other. It is possible to retrieve the emission angle $\alpha'_{3,F}$ from the conservation of the Poynting vector:

$$k_1^\omega \sin(\alpha'_1) + k_2^\omega \sin(\alpha'_2) = k_{3,F}^{2\omega} \sin(\alpha'_{3,F}) \qquad (1)$$

While $\alpha'_{3,B}$ is obtained from the vectorial sum of the two pumps' wavectors, in the three waves interaction:

$$\vec{k}_{1,}^\omega + \vec{k}_2^\omega = \vec{k}_{3,B}^{2\omega} \qquad (2)$$

The obtained propagation angle of the noncollinear bound wave, $\alpha'_{3,B}$, is represented in Fig. 3, and results as a combination of the propagation angles of the two pump beams.

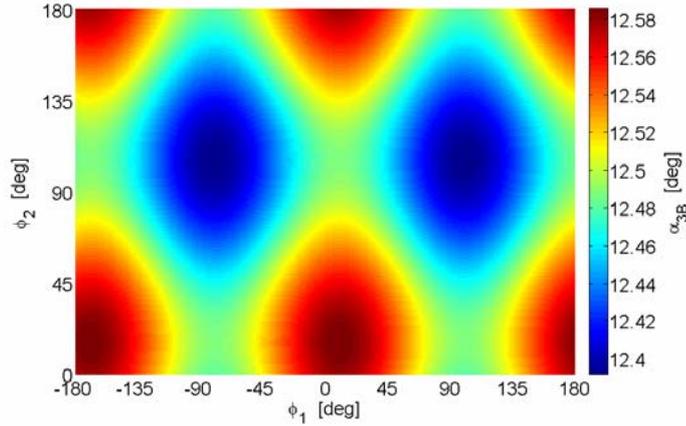

Fig. 3. Propagation angle for the noncollinear bound wave, $\alpha'_{3,B}$ as a function of polarization state of both pump beams.

It is interesting to identify some requirements for the noncollinear bound and free waves to interfere with each other inside the nonlinear material, by discussing the pumps' superposition, the effect of anisotropy and the interaction length. First of all, the noncollinear bound wave is traveling along with the two pump beams superposition, thus it survives as long as the two pump profiles are overlapping within the sample length. The material birefringence may determine, or increase the spatial separation between the noncollinear bound and free waves within the crystal. This effect in turn can be avoided using reduced incidence angles as well as high refractive index materials, thus constraining the internal propagation angles, $\alpha'_1$ and $\alpha'_2$, to limited values and implying small difference between the indices seen by the two waves. Finally, pulse duration, $\tau_P$, and material dispersion play a significant role in this process, since they may affect the effective interaction length. The temporal walk-off, arising from group velocity dispersion, must be longer than crystal length, in order to let the bound and the free waves interfere.

**3. Noncollinear second harmonic generation.**

The full expression of the SH power in the noncollinear scheme, $P_{\omega1+\omega2}$, as a function of incidence angles, including the effect of absorption, trough the extinction coefficient at the fundamental, $k_\omega$, and at the second harmonic frequency, $k_{2\omega}$, can be written as:

$$P_{\omega1+\omega2}(\alpha) = \left(\frac{512\pi^3}{A_1 A_2}\right)(t_{\omega1})^2 \cdot (t_{\omega2})^2 \cdot T_{\omega1+\omega1} \cdot P(\alpha_1,\alpha_2) \cdot \frac{\sin^2(\Psi^{HH}(\alpha)) + \sinh^2(\chi^{HH}(\alpha))}{\Psi^{HH}(\alpha)^2 + \chi^{HH}(\alpha)^2} \left(\frac{\pi L}{\lambda}\right)^2 (d_{eff}(\alpha))^2 \quad (3)$$

where $A_1$ and $A_2$ are the fundamental beams transverse areas onto sample surface, retrieved from the pump beam area (A) as $A_i = A/cos(\alpha_i)$, while $t_{\omega1}(\alpha_1,\phi_1)$ and $t_{\omega2}(\alpha_2,\phi_2)$ are the Fresnel transmission coefficients for the fundamental fields at the input interface, $T_{\omega1+\omega2}(\alpha,\phi)$ is the Fresnel transmission coefficient for the SH power at the output interface, and $L$ is sample thickness. It's worth noting that Fresnel coefficients are in general complex, but for small extinction coefficients they can be assumed real. The power of the incident fundamental beams is taken into account in the term $P(\alpha_1,\alpha_2) = P_{\omega1} \cdot P_{\omega2} \cdot e^{-2(\delta_1+\delta_2)}$, which also includes an attenuation factor given by:

$$\delta_1 + \delta_2 = \left(\frac{\pi L}{2}\right)\frac{2}{\lambda}\left[k_\omega \cos(\alpha'_1) + k_\omega \cos(\alpha'_2) + 2k_{2\omega} \cos(\alpha'_{3,F})\right] \quad (4)$$

The phase factor, $\Psi^{HH}(\alpha)$, in the noncollinear scheme is given by:

$$\Psi^{HH}(\alpha) = \left(\frac{\pi L}{2}\right)\left(\frac{2}{\lambda}\right)\left[n_\omega(\alpha'_1) \cdot \cos(\alpha'_1) + n_\omega(\alpha'_2) \cdot \cos(\alpha'_2) - 2n_{2\omega}(\alpha'_3) \cdot \cos(\alpha'_3)\right] \quad (5)$$

Due to materials' anisotropy, $n_\omega(\alpha'_1)$, $n_\omega(\alpha'_2)$ and $n_{2\omega}(\alpha'_3)$ depend on the polarization angles $\phi_1$, $\phi_2$ and $\phi_3$ as well.

The term $\chi^{HH}(\alpha)$ introduced by Herman and Hayden is an additional phase term given by the imaginary part of the refractive index, which takes into account the effect of absorption on the Maker fringes. In fact, as well as the material absorption reduces the amplitude of the wave, the shape of the angular curve itself is modified by [23]:

$$\chi^{HH}(\alpha) = \left(\frac{\pi L}{2}\right)\left(\frac{2}{\lambda}\right)\left[k_\omega \cdot \cos(\alpha'_1) + k_\omega \cdot \cos(\alpha'_2) - 2k_{2\omega} \cdot \cos(\alpha'_3)\right] \quad (6)$$

Finally, the term $d_{eff}(\alpha)$ in Equation (3) represents the effective susceptibility tensor, being dependent on the second order nonlinear optical tensor, the polarization state of both pumps and generated beams and, of course, on the fundamental beams incidence angles, $\alpha_1$ and $\alpha_2$.

**4. Examples for some different crystal structures.**

In the examples that follow, we will investigate noncollinear SHG from two crystal structures, characterized by different type non-zero nonlinear optical susceptibility. We chose two common nonlinear optical structures, without loosing generality. These examples can be extended to any nonlinear optical structure, provided the symmetry of the nonlinear optical tensor is known. The two incident angles were fixed to $\alpha_1 = 26°$ and $\alpha_2 = 44°$, with respect to surface normal, in both examples. The generated beam is projected along the bound SH angle, i.e. approximately along the bisector of the two pump beams aperture angle.

The first example is the symmetry group 6mm, corresponding to Gallium and Aluminum nitrides (GaN, AlN) and their alloys and several semiconductors as ZnO, CdS and Cd Se or SiC, to name some nonlinear optical materials. To fix our attention on one particular case, we consider a SiC slab 50 μm thick, i.e. several coherence lengths, in order to visualize in the resulting polarization chart several fringes due to the interference between the bound and the free waves, and to see the effect of increasing absorption on the fringes. Given the two incident fields as $\vec{E}_i = (sin(\phi_i) \quad -cos(\phi_i)cos(\alpha_i) \quad -cos(\phi_i)sin(\alpha_i))$, since the $\tilde{d}$ tensor corresponding to SiC has only three independent nonzero-components, $d_{15}=d_{24}$, $d_{31}=d_{32}$ and $d_{33}$, the final expressions for $d_{eff}(\alpha)$ as a function of polarization angle of the two pumps, $\phi_1$ and $\phi_2$ is easily written:

$$d_{eff}^{\hat{s}} = -d_{15}\left[sin(\phi_1)cos(\phi_2)sin(\alpha'_2) + cos(\phi_1)sin(\phi_2)sin(\alpha'_1)\right]$$

$$d_{eff}^{\hat{p}} = -d_{24}cos(\alpha'_3)\left[cos(\alpha'_1)sin(\alpha'_2) + sin(\alpha'_1)cos(\alpha'_2)\right]cos(\phi_1)cos(\phi_2) -$$
$$- sin(\alpha'_3)\left\{d_{31}sin(\phi_1)sin(\phi_2) + cos(\phi_1)cos(\phi_2)\left[d_{32}cos(\alpha'_1)cos(\alpha'_2) + d_{33}sin(\alpha'_1)sin(\alpha'_2)\right]\right\}$$
(7)

where the apex $\hat{s}$ or $\hat{p}$ stands for the polarization state of the generated beam.

We assumed the same fundamental wavelength, $\lambda_\omega=830$, and the linear refractive indices already used for the calculations shown in Fig.2 and Fig. 3 [22]. The SH power for $\hat{p}$- and $\hat{s}$-polarization state, was analytically calculated as a function of the polarization state of both fundamental beam, i.e. by systematically varying $\phi_1$ and $\phi_2$.

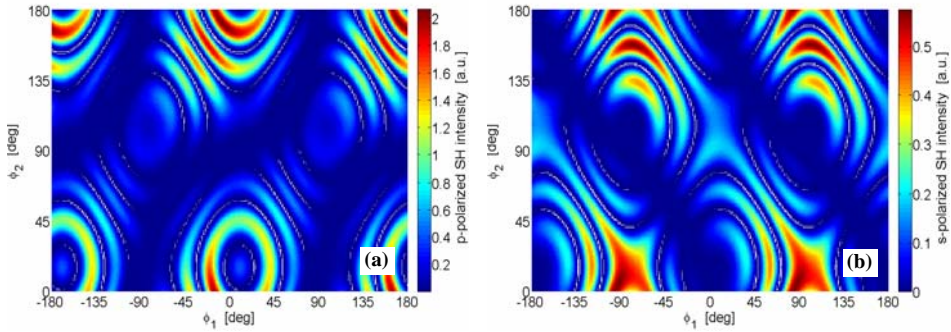

Fig.4. Second harmonic intensity as a function of the polarization state of the first pump beam ($\phi_1$) and the second pump beam ($\phi_2$), calculated for a 50 μm SiC (crystal structure 6mm) slab without absorption. The polarization state of the analyzer is set to (a) $\hat{p}$ - and (b) $\hat{s}$ -, respectively.

In Fig.4 we show the calculated SH intensity obtained when the nonlinear material is non absorbing, i.e. $k_{2\omega}=0$. The oscillation of SH signal as a function of different pump beams polarization state is due to the high thickness, with respect to the coherence length of the process. At the given incidence angles $\alpha_1$ and $\alpha_2$, in fact, the coherence length, defined in noncollinear SHG as $l_c = \dfrac{\pi c}{\omega \left| n_{\omega 1}(\alpha'_1) \cdot cos(\alpha'_1) + n_{\omega 2}(\alpha'_2) \cdot cos(\alpha'_2) - 2n_{2\omega}(\alpha'_3) \cdot cos(\alpha'_3) \right|}$, ranges between its maximum value of $l_c$=1.75 μm, at $\phi_1$=90° and $\phi_2$=90°, and its minimum value at $\phi_1$=0° and $\phi_2$=0° ($l_c$=1.35 μm).

We now consider a variety of absorption condition, that may be obtained for instance by ion-implantation during the fabrication of the nonlinear optical crystal [24]. By introducing the effect of linear optical absorption at the SH frequency, we observe the double effect of attenuation of the absolute value of the generated signal and the reduced contrast in the interference fringes pattern. In Fig.5 we show the calculated SH power, for $\hat{p}$- and $\hat{s}$- polarization state, obtained for an absorption coefficient at 2ω of $k_{2\omega}$=0.003, corresponding to a linear transmittance at $\lambda_{2\omega}$=415nm of approximately 10%.

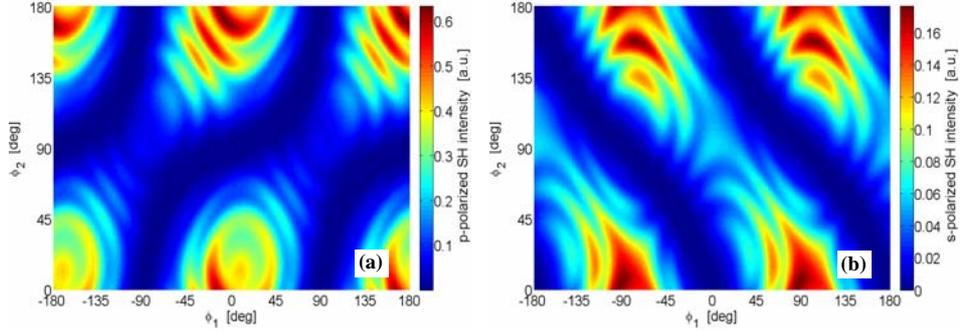

Fig.5. Second harmonic intensity as a function of the polarization state of the first pump beam ($\phi_1$) and the second pump beam ($\phi_2$), calculated for a 50 μm SiC slab for an absorption coefficient $k_{2\omega}$=0.003 at 2ω. The polarization state of the analyzer is set to (a) $\hat{p}$- and (b) $\hat{s}$-, respectively.

A more interesting case develops when the linear optical absorption at 2ω is further increased. The only surviving wave is the bound wave, since it experiences the same refractive index and linear absorption of the two pump waves, even thought the absorption coefficient at 2ω is high. On the other side, the free wave is generated and then immediately absorbed within the nonlinear material, thus we observe the disappearance of the fringes in the SH signal. From the calculation shown in Fig.6 we can appreciate this effect, when the absorption coefficient at 2ω is assumed to be $k_{2\omega}$=0.005, corresponding to a linear transmittance at 2ω of approximately 2%.

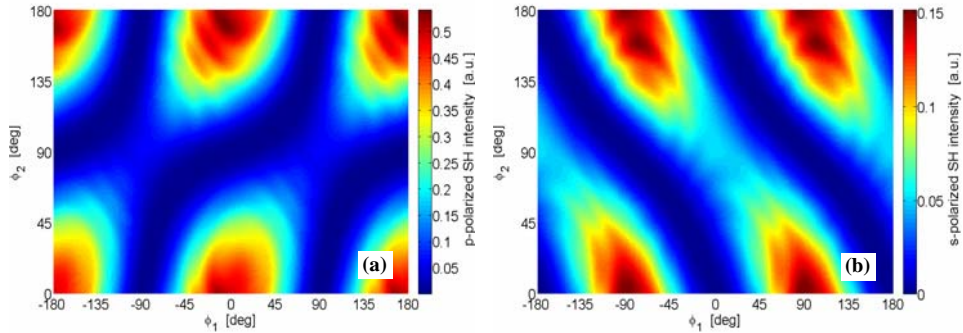

Fig.6. Second harmonic intensity as a function of the polarization state of the first pump beam ($\phi_1$) and the second pump beam ($\phi_2$), calculated for a 50 μm SiC slab for an absorption coefficient $k_{2\omega}$=0.005 at 2ω. The polarization state of the analyzer is set to (a) $\hat{p}$- and (b) $\hat{s}$-, respectively.

From Fig.6 it is also possible to make some consideration on the pattern of the polarization map. According to Equations (7), the noncollinear SH signal generated in $\hat{p}$ polarization (Fig.6a), shows that the absolute maxima are achievable when both pumps are $\hat{p}$-polarized, i.e. when $\phi_1$ and $\phi_2$ are both 0° or 180°, while relative maxima (saddle point) occur when both pumps are $\hat{s}$-polarized, i.e. when polarization angles of both pumps are set to ± 90°. Conversely, when the two pump beams have crossed polarization, i.e when $\phi_1 = 0°$ and $\phi_2 = 90°$ and viceversa, the nonlinear optical tensor do not allow SH signal which is $\hat{p}$-polarized. On the other hand, when the analyzer is set to $\hat{s}$-polarization, the maxima generally occur when the two pump beams have crossed polarization, as shown in Figure 6b, i.e. when the first pump is $\hat{s}$-polarized and the second pump is $\hat{p}$-polarized, i.e. $\phi_1 = \pm 90°$ and $\phi_2$ is equal to either 0° or 180°. Relative maxima occur in the reverse situation, when the first pump is $\hat{p}$-polarized, $\phi_1 = 0°$ or $\pm 180°$, and the second pump $\hat{s}$-polarized, $\phi_2 = 90°$. Finally, when the two pumps are equally polarized, either $\hat{s}$ or $\hat{p}$, there is no SH signal $\hat{s}$-polarized.

We then investigated the nonlinear optical response of a different crystal structure, 43m, corresponding to a cubic cell, which is characteristic of GaAs, GaP, InAs, InP. A 50 μm thick GaAs slab was chosen as nonlinear material. In this case we tune the fundamental wavelength to $\lambda_\omega$=1500 nm so that the generated beam falls within the absorption band of GaAs ($E_g \sim 1.42$ eV). The parameters provided by refractive index dispersion are $n_\omega$=3.37 at the fundamental beam frequency and $n_{2\omega}$=3.70 at the SH frequency, without birefringence, along with a high absorption coefficient such that at 2ω $k_{2\omega}$=0.1 [25]. Due to its high-symmetry nonlinear-optical susceptibility tensor, the only non-zero elements are $d_{14}= d_{25}= d_{36}$. The expressions for $d_{eff}(\alpha)$ as a function of polarization angle of the two pumps, $\phi_1$ and $\phi_2$ is thus simplified:

$$d_{eff}^{\hat{s}} = d_{14} cos(\phi_1) cos(\phi_2) \left[ cos(\alpha'_1) sin(\alpha'_2) + sin(\alpha'_1) cos(\alpha'_2) \right]$$
$$d_{eff}^{\hat{p}} = cos(\alpha'_3) d_{25} \{ sin(\phi_1) cos(\phi_2) sin(\alpha'_2) + sin(\phi_2) cos(\phi_1) sin(\alpha'_1) \} + \qquad (8)$$
$$+ sin(\alpha'_3) d_{36} \{ sin(\phi_1) cos(\phi_2) cos(\alpha'_2) + sin(\phi_2) cos(\phi_1) cos(\alpha'_1) \}$$

The calculated noncollinear SH signal, for the $\hat{p}$- and $\hat{s}$-polarization state, is depicted in Fig.7. Although the relative and absolute SH maxima appear under the same polarization conditions, the figure suggests a different pattern in the $\hat{p}$-polarized curve, with respect to the crystal structure 6mm, due to the different dependence of the $d_{eff}(\alpha)$. The pattern of the noncollinear SH signal, in fact, can be modified depending on the different crystalline structure that is considered.

Furthermore, the high absorption coefficient is responsible for the full disappearance of the noncollinear free wave thus the curves appear completely smoothed and there is no trace of fringes. Meanwhile, the fundamental beam, tuned to a range of optical transparency, imposes its propagation properties on the bound noncollinear wave which is made able to propagate without being affected by absorption.

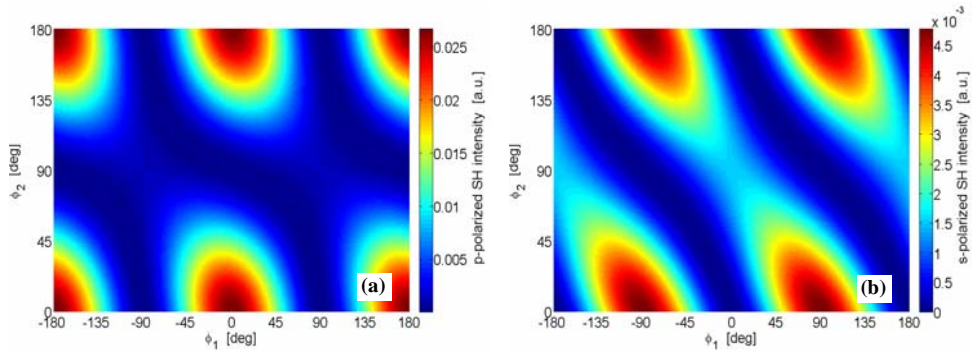

Fig.7. Second harmonic intensity as a function of the polarization state of the first pump beam ($\phi_1$) and the second pump beam ($\phi_2$), calculated for a 50 μm GaAs slab (crystal structure 43m) under high SH absorption, i.e. $k_{2\omega}$=0.1 at 2ω. The polarization state of the analyzer is set to (a) $\hat{p}$ - and (b) $\hat{s}$ -, respectively.

Finally, by tuning the fundamental frequency around the band edge it is possible to investigate the effect of different levels of absorption, and to observe the occurrence of the interference fringes as long as the second harmonic frequency is approaching the transmission band.

## 5. Conclusions.

We have theoretically investigated noncollinear SHG in crystal structures some coherence length thick. We shown that the interference between the bound and the free noncollinear waves can be evidenced by changing the polarization state of both fundamental beams, while the incidence angle is fixed. The effect of linear optical absorption at 2ω, thus on the free wave propagation, was also investigated and we demonstrated how the pattern of the noncollinear SH signal is modified, when ranging from an absorption-free to a highly absorptive crystal.

## Acknowledgments.

Stefano Paoloni, Marco Centini and Alessio Benedetti are kindly acknowledged for interesting discussions.

## References and links